\documentclass{sig-alt-release2}
\usepackage{amsmath}
\usepackage{amssymb}
\usepackage{amsfonts}
\usepackage[overload]{textcase}
\usepackage{xspace}
\usepackage{graphics}

\newcommand{\eqpunc}[1]{\,{#1}}
\newcommand{\mydef}{\stackrel{{\rm def}}{=}}

\newcommand{\chaabbrev}{CH}
\newcommand{\chao}{\mbox{\chaabbrev-1}\xspace}
\newcommand{\chat}{\mbox{\chaabbrev-2}\xspace}
\newcommand{\sj}{Sp\"arck Jones\xspace}

\newcommand{\pmshort}{RSJ-PM\xspace}
\newcommand{\rwshort}{RW\xspace}

\newcommand{\corpussize}{N}

\newcommand{\docfreq}{n_{\attrind}}

\newcommand{\attrvar}{X}

\newcommand{\attrind}{i}

\newcommand{\attrvarind}{\attrvar_\attrind}

\newcommand{\relvar}{R}
\newcommand{\rely}{{\tt y}}
\newcommand{\reln}{{\tt n}}
\newcommand{\rel}{\relvar = \rely}
\newcommand{\nonrel}{\relvar = \reln}

\newcommand{\est}[1]{\widehat{#1}}
\newcommand{\probsym}{P}

\newcommand{\condprobgen}[3]{#1(#2\,|\,#3)}
\newcommand{\condprob}[2]{\condprobgen{\probsym}{#1}{#2}}

\newcommand{\ogr}{p_\attrind} 
\newcommand{\ognr}{q_\attrind}

\newcommand{\ogrch}{\est{\ogr^{{\rm CH}}}}
\newcommand{\ognrch}{\est{\ognr^{{\rm CH}}}}
\newcommand{\pprobconst}{\pi}
\newcommand{\pprobconstrw}{\pprobconst}
\newcommand{\pratioconst}{\pprobconst'}
\newcommand{\ogrrw}{\est{\ogr^{{\rm RW}}}}
\newcommand{\newest}{\est{\ogr}}

\newcommand{\boostconst}{L}
\newcommand{\boostfn}{\boostconst(\docfreq)}

\newcommand{\theboost}{lift\xspace}
\newcommand{\theboostalt}{boost\xspace}

\DeclareFixedFont{\auacc}{OT1}{phv}{b}{n}{18}

\begin{document}
\conferenceinfo{SIGIR'07,} {July 23--27, 2007, Amsterdam, The Netherlands.}
\CopyrightYear{2007}
\crdata{978-1-59593-597-7/07/0007}

\newcommand{\newcite}[1]{\cite{#1}}
\newcommand{\twocite}[2]{(\cite{#2}, #1)}
\newcommand{\twonewcite}[2]{(\cite{#2}, #1)}

\title{IDF  Revisited: A  Simple New Derivation within the
  Robertson-Sp{\auacc \"a}rck
  Jones  Probabilistic  Model}
\numberofauthors{1}
\author{Lillian Lee \\ \affaddr{Dept. of Computer Science, Cornell
    University} \\ \affaddr{Ithaca, NY 14853-7501 USA} \\ \affaddr{http://www.cs.cornell.edu/home/llee} \\ \email{llee@cs.cornell.edu}}
\maketitle

\begin{abstract}
There have been a number of prior attempts to theoretically justify the effectiveness of the
inverse document frequency (IDF).  Those that take as their starting
point Robertson and Sp\"arck Jones's probabilistic model are based on
strong or complex assumptions.
We show that a more intuitively plausible assumption suffices.
Moreover, the new assumption, while conceptually very simple, provides a
solution to an estimation problem that had been deemed intractable by
Robertson and Walker (1997).
\end{abstract}

\vspace{1mm}
\noindent
{\bf Categories and Subject Descriptors:} H.3.3 {[Information Search and Retrieval]}: {Retrieval models}

\vspace{1mm}
\noindent
{\bf General Terms:} Theory, Algorithms

\vspace{1mm}
\noindent
{\bf Keywords:} inverse document frequency, IDF, probabilistic model,
term weighting

\section{Introduction}

The inverse document frequency (IDF) \cite{SparckJones:72a} 
has been ``incorporated in (probably) all
information retrieval systems'' \twocite{pg. 77}{Harman:05a}.
Attempts to theoretically explain its empirical
successes  abound
\twocite{{\it inter alia}}{Croft+Harper:79a,Wong+Yao:89a,Church+Gale:95a,Robertson+Walker:97a,Greiff:98a,Papineni:01a,Fang+Tao+Zhai:04a,deVries+Roelleke:05a}.
Our focus here is on explanations based on Robertson and \sj's
\emph{probabilistic-model} (\pmshort) paradigm of information
retrieval \cite{Robertson+SparckJones:76a}, 
not because of 
any prejudice against other paradigms, 
but because a certain {\pmshort}-based justification of the IDF in the
absence of relevance information has been
promulgated by several influential authors \cite{Croft+Harper:79a,Robertson:04a,Manning+Raghavan+Schuetze:priv:06a}.

\pmshort-based accounts use either an assumption due to Croft and
Harper \cite{Croft+Harper:79a} that is mathematically
convenient but not plausible in real settings, or a complex assumption
due to Robertson and Walker \cite{Robertson+Walker:97a}.  We show that
the IDF can be derived within the \pmshort framework via a new
assumption that directly instantiates a highly intuitive notion, and
that, while conceptually simple, solves an estimation problem
deemed intractable by Robertson and Walker
\cite{Robertson+Walker:97a}.

\section{Croft-Harper derivation}

In the (binary-independence version of the) \pmshort,  the $\attrind^{th}$ term is assigned weight
\begin{equation}
 \log \frac{\ogr (1 - \ognr)}{\ognr (1 - \ogr)}\eqpunc{.}
\label{eq:rsj}
\end{equation}
where  $\ogr \mydef
\condprob{\attrvarind=1}{\rel}$, $\ognr \mydef \condprob{\attrvarind =
  1}{\nonrel}$, $\attrvarind$ is an indicator variable for the presence of the $\attrind^{th}$ term, and
$\relvar$ is a relevance random variable.
Croft and
Harper \cite{Croft+Harper:79a} proposed the use of two assumptions to estimate $\ogr$ and
$\ognr$ in the absence of relevance information.
{\bf \chao}, which is unobjectionable, simply states that most of the documents
in the corpus are not relevant to the query.
This allows us to set
$\ognrch \mydef \frac{\docfreq}{\corpussize}\eqpunc{,}$
where $\docfreq$ is the number of documents in the corpus that
contain the $\attrind^{th}$ term, and $\corpussize$ is the number of
documents in the corpus.  
The second assumption, {\bf \chat}, is that {all query terms 
share the same
 probability $\pprobconst$ of occurring in a relevant
 document\footnote{This can be relaxed to apply to just the query terms in the
 document in question.}.}
Under \chat, one sets $\ogrch \mydef
\pprobconst$, and thus (\ref{eq:rsj}) becomes
\begin{equation}
\pratioconst + \log \frac{\corpussize -
  \docfreq}{\docfreq}\eqpunc{,}
\label{eq:chmatch}
\end{equation}
where $\pratioconst = \log\left(\pprobconst/(1 - \pprobconst)\right)$
is constant (and is 0 if $\pprobconst=.5$). Quantity (\ref{eq:chmatch}) is
essentially the IDF.

  \chat is an ingenious device for pushing the
derivation above through.
However, 
intuition suggests
 that the occurrence
probability of query terms in relevant documents should be at least
somewhat correlated with their occurrence probability in arbitrary
documents within the corpus, and hence not constant.  For example, a
very frequent term can be
expected to occur in a noticeably large fraction of any particular subset of the
corpus, including the relevant documents.  Contrariwise, a query term
might be relatively infrequent overall due to having a more commonly
used synonym; such a term would still occur relatively infrequently
even  within the set of (truly) relevant documents.\footnote{Indeed,
one study  \cite{Greiff:98a} did find $\ogr$  increasing with
 $\docfreq$.}

\section{Robertson-Walker derivation} Robertson and Walker (\rwshort) \cite{Robertson+Walker:97a} 
also object to \chat, on the grounds that for query terms with very large document frequencies, weight (\ref{eq:chmatch}) can be
negative.
This anomaly, they show,
arises precisely because $\ogrch$ is  constant.
They then propose  the following alternative:
\begin{equation*}
\ogrrw  \mydef   \frac{\pprobconst}{\pprobconstrw +
 (1 - \pprobconstrw) \frac{\corpussize-\docfreq}{\corpussize}}\eqpunc{,}
\label{eq:rwa}
\end{equation*}
where $\pprobconstrw$ is the Croft-Harper constant, but
reinterpreted as the estimate for $\ogr$ just when $\docfreq=0$.
One can 
check that $\ogrrw \in [\pprobconst,1]$ slopes up
hyperbolically in $\docfreq$. Applying $\ogrrw$ and
$\ognrch$ to the term-weight scheme (\ref{eq:rsj}) yields 
\begin{equation}
\pratioconst + \log \frac{\corpussize}{\docfreq}\eqpunc{ }
\label{eq:rwmatch}
\end{equation}
(which is  positive as long as $\pprobconst \geq .5$).  

\section{New assumption}
\label{sec:new}

The estimate $\ogrrw$  increases monotonically in
$\docfreq$, which is a desirable property, as we have argued above.  However, its exact functional form
does not seem particularly intuitive.  
\rwshort motivate it simply as an approximation
to a linear form; approximation is necessary, they claim, because
\begin{quote}
the
straight-line model [i.e., $\ogr$ linear in $\ognr$ and hence $\docfreq$ by {\chao}] is actually rather intractable, and does not lead
to a simple weighting formula
\twocite{pg. 18}{Robertson+Walker:97a}.\footnote{The fact that
  \rwshort's Figure
  2  depicts the linear scenario graphically
appears to have led to some mistaken impressions (e.g.,
\cite{Greiff:98a}, pg. 18, coincidentally) that this is the
mathematical model that \rwshort actually employed.}  
\end{quote}
{Despite this claim, we show  here that there exists a highly
  intuitive linear estimate
  that leads to a term weight varying inversely with document frequency.}

There are two main principles that motivate our new estimate.  First,
as already stated, any estimate of $\ogr$ should be  positively correlated
with $\docfreq$.  The second and key insight is that \emph{query
terms should 
have a higher occurrence probability within relevant
documents than within the document collection as a whole}.  Thus, if
the $\attrind^{th}$ term appears in the query, we
should ``\theboost'' its estimated occurrence
probability in relevant documents above $\docfreq/\corpussize$, which
is its estimated occurrence
probability in general documents.  This leads us to the following
intuitive estimate, which is reminiscent of ``add-one smoothing''
used in language modeling (more on this below):
\begin{equation}
\newest \mydef \frac{\docfreq + \boostconst}{\corpussize +
  \boostconst}\eqpunc{.} 
\label{eq:newest}
\end{equation}
Here the
$\boostconst > 0$ in the numerator\footnote{The $\boostconst$ in the denominator ensures that
  $\newest \leq 1$.} is a
``\theboost'' or ``\theboostalt'' constant.%
\footnote{Since the
  \pmshort document-scoring 
  function  only accumulates weights for
 terms appearing in the query, it is fine to compute the $\newest$'s
  offline, that is, before the query is seen.}  
Plugging  $\newest$ and $\ognrch$ into (\ref{eq:rsj}) yields the term weight
\newcommand{\mypi}{\frac{\docfreq  + \boostconst}{\corpussize + \boostconst}}
\newcommand{\myqi}{\frac{\docfreq}{\corpussize}}
\newcommand{\myqicomp}{\frac{\corpussize - \docfreq}{\corpussize}}
\newcommand{\mypicomp}{\frac{\corpussize - \docfreq}{\corpussize + \boostconst}}
\begin{align*}
\log  \bigg(  \frac{\mypi}{\myqi}  
 \frac{\myqicomp}{\mypicomp}\bigg) 
 &=  \log \left(1 + \frac{\boostconst}{\docfreq}\right) \eqpunc{,}
\end{align*}
which varies inversely in $\docfreq$, as desired.

Furthermore, as hinted at above,  selecting $\boostconst$'s value
is equivalent to selecting $\newest$'s value for query terms whose
document frequency is 0.  That is,  $\boostconst/(\corpussize +
\boostconst)$ is directly analogous to $\pprobconst$ in 
\rwshort's derivation.  Indeed, choosing $\boostconst = \corpussize$
is just like choosing $\pprobconst = 0.5$, which is commonly done in
presentations of the Croft-Harper
derivation in order to eliminate
the leading constant $\pratioconst$ in (\ref{eq:chmatch}); doing so in
our case yields the following term weight, which is the ``usual''
form of the IDF
\twocite{pg. 184}{Witten+Moffat+Bell:99a}:
\[ \log \left(1 +
\frac{\corpussize}{\docfreq}\right)\eqpunc{.}
\]
Finally, note that 
$\newest$ is linear in $\docfreq$; we have thus contradicted the assertion
 quoted above
 that developing a ``straight-line'' model is ``intractable'' \cite{Robertson+Walker:97a}.

\section{Onward and upward} 
An interesting direction for future work is to consider
\theboost\xspace {\em functions} $\boostfn$ that depend on $\docfreq$.  It can
be shown that different choices of $\boostfn$ allow one to model
{\em non-linear} dependencies of $\ogr$ on $\docfreq$ that occur
in real data, such as the 
approximately logarithmic
dependence observed in TREC corpora by Greiff \shortcite{Greiff:98a}.  Importantly, seemingly similar choices of
$\boostfn$ yield strikingly different term-weighting schemes;
it would be interesting to empirically compare these new schemes
against the classic IDF.

\vspace*{-2mm}

\paragraph*{{\bf Acknowledgments}} We thank Jon Kleinberg and the anonymous
reviewers for  helpful comments. This paper is based upon work
supported in part by the National Science Foundation under grant
no. IIS-0329064, a Yahoo! Research Alliance gift, and an Alfred
P. Sloan Research Fellowship. Any opinions, findings, and conclusions
or recommendations expressed are those of the author and do not
necessarily reflect the views or official policies, either expressed
or implied, of any sponsoring institutions, the U.S. government, or
any other entity.

\scriptsize

\vspace*{-.8mm}
\bibliographystyle{abbrv}

\end{document}